\documentclass[12pt,fleqn]{article}
\input{epsf}
\usepackage{amsmath}
\usepackage{latexsym}
\usepackage{amssymb}
\newlength{\dinwidth}
\newlength{\dinmargin}
\newcommand{\resection}[1]{\setcounter{equation}{0}\section{#1}}

\newcommand{\f}[2]{\frac{#1}{#2}}

\newcommand{\nn}{\nonumber}

\def\kh{{\hat k}}
\def\t{\tau}
\def\d{\partial}
\def\l{\lambda}
\def\dk{\f{d^6k}{(2\pi)^6}}
\def\pts{(\phi^3)_6}
\def\xb{{\bar x}}
\def\as{\alpha_S}
\def\xp{x_{I\!\!P}}       
\def\hs{\hspace{1mm}}
\def\hsp{\hspace{0.5mm}}
\begin{document}
 \thispagestyle{empty}

\begin{flushright}
 CERN-TH/97-248\\
 UPRF-97-011\\
 September 1997\\
\end{flushright}
\vspace*{1 cm}
 \begin{center}
 {\huge  Fracture Functions from Cut Vertices}\\
\vskip 1.0cm
\begin{Large}
{M.~Grazzini$^{a}$, L.~Trentadue$^{a}$ and G.~Veneziano$^{b}$}\\
\end{Large}
\vskip .5cm
{$^a$ \it Dipartimento di Fisica, Universit\`a di Parma and\\
 INFN Gruppo Collegato di Parma, 43100 Parma, Italy\\
 $^b$ \it Theory Division, CERN, CH-1211, Geneva, Switzerland}\\
\end{center}

 \begin{abstract}
 Using a generalized cut vertex expansion we
introduce the concept of an extended fracture
 function for the description of semi-inclusive
 deep inelastic processes in the target fragmentation region.
 Extended fracture functions are shown to obey
 a standard homogeneous DGLAP-type equation
which, upon integration over $t$, becomes
 the usual inhomogeneous evolution equation for ordinary fracture functions.
 \end{abstract}
\vspace*{2cm}
\centerline{PACS 13.85.Ni}

\newpage
\setcounter{page}{1}

\resection{Introduction}

Within QCD, Wilson's Operator Product Expansion (OPE) \cite{wilson}
finds its most straightforward application in lepton hadron
Deep Inelastic Scattering (DIS). The cross section
for this process is related to
the time-ordered product of two currents,
$T[J(x)J(0)]$, which, at short distances, is expanded as
\begin{equation}
T[J(x)J(0)]=\sum_k C_k(x)\hs O_k(0).
\end{equation}
The coefficient functions $C_k(x)$ embody the short-distance
structure of the
amplitude, while $O_k$ are local operators whose matrix elements describe
long-distance effects.
For DIS the OPE  leads to an  expansion of the forward
Compton amplitude
which, by use of dispersion relations \cite{christ},
 gives  predictions for the moments of the DIS inclusive cross section.

Unfortunately,
for more general hard processes the straightforward use
of the OPE technique fails. Two ways out of this impasse have been
proposed. The first one \cite{factorization} relies  on a careful
study of collinear and soft singularities order by order in
perturbation theory
and aims at proving that such singularities can  all be lumped into
some universal (i.e.
hard-process independent) functions. This method
has the advantage of being close to the physical parton picture for the
hard process. As an alternative,
an expansion in terms of {\em cut vertices} was proposed by Mueller
\cite{mue} as a more direct extension  of the OPE. Although more formal,
 the method of cut vertices has
the advantage of giving directly
an expansion for the absorptive part of the amplitude, i.e. for the
cross section, in terms of the convolution of a cut vertex with
 a coefficient function.
The cut vertex contains all the long-distance effects,
and, as such,  generalizes the
 matrix element of a local operator,
while the coefficient function, as usual, is
calculable in perturbation theory.
The cut vertex method
has been used in Ref.\cite{gm} to deal with a variety
of hard processes in QCD.

Let us consider a deep inelastic semi-inclusive reaction
in which  a hadron in the  final state is detected.
When the transverse momentum of the hadron is of order $Q$ the usual
current fragmentation mechanism holds \cite{aemp}.
In the region in which the transverse
momentum of the produced hadron is much smaller than $Q$,
the so-called target fragmentation region,
 the simple description based on parton densities
and fragmentation functions fails.
 For this reason, in order to describe
the target fragmentation region as well,
 a new formulation of semi-inclusive hard processes in terms of  {\em  
fracture functions}
was proposed \cite{frac} and later developed \cite{grau,altri}.
A fracture function $M^i_{AA^\prime}(x,z,Q^2)$ gives the
 probability of extracting a parton $i$ with momentum fraction
$x$ from a hadron $A$ while
 observing a final hadron $A^\prime$ in the target fragmentation region
with longitudinal momentum fraction $z$. The possible relevance
of such an object was already mentioned by Feynman \cite{Feynman}
before the advent of QCD.

In this paper we will consider the case
in which  also the transverse momentum of the hadron $A^\prime$
 (or equivalently the momentum transfer $t=-(p_A-p_{A^\prime})^2$) is  
observed.
We will argue that, in the region $t\ll Q^2$, an expansion
in terms of cut vertices can be given and that a $t$ dependent (or extended)
fracture function\footnote{Similar objects have been defined in the context
of diffraction in Ref. \cite{soper}.}
can be defined  in terms of a new kind of cut vertex.
We will show, within a $\pts$ toy-model field theory,
that extended factorization is just a consequence of
 collinear power counting \cite{sterman,css}.

The paper is organized as follows. In Sect. 2 we recall the
definition of cut vertices and we obtain the expansion of the moments  
of the structure function
in terms of matrix elements of local operators times coefficient functions
by using cut vertices. In Sect. 3 the cut vertex approach is extended  
to deal
with semi-inclusive processes and a generalized cut vertex expansion  
is given.
In Sect. 4 we define extended fracture functions and their evolution  
equation
is given and discussed.
In Sect. 5 we discuss the results and draw our conclusions.

\resection{Cut vertices}

In this section we will recall
the definition of cut vertices in the simpler case of  $\pts$   theory.
This toy model, despite  its  simpler structure,
shares  several important properties with QCD. It has a dimensionless,  
asymptotically free coupling constant and the diagrams with leading
mass singularities
have the same topology as
in (light cone gauge) QCD. For these reasons  $\pts$
is an excellent theoretical framework for the study of
factorization
properties \cite{phicubo}.

Consider the inclusive deep inelastic process
\begin{equation}
p+J(q)\rightarrow X\nn
\end{equation}
off the current $J=\f{1}{2} \phi^2$.
We define as usual $Q^2$ and $x$ as
\begin{equation}
Q^2=-q^2~~~~~~x=\f{Q^2}{2pq} \; .
\end{equation}
Let us choose a frame in which $p=(p_+,p_-,{\bf 0})$ with $p_+\gg p_-$
and $pq\simeq p_+q_-$.
Given a vector $k=(k_+,k_-,{\bf k})$ define $\kh=(k_+,0,{\bf0})$.
The structure function, defined as
\begin{equation}
F(p,q)=\f{Q^2}{2\pi} \int  d^6y~e^{iqy} <p|J(y)J(0)|p>,
\end{equation}
 describes the interaction of the far off-shell current $J(q)$ with
an elementary quantum of momentum $p$
through
the discontinuity
of the forward scattering amplitude (see Fig. 1).

\begin{figure}[htb]
\begin{center}
\begin{tabular}{c}
\epsfxsize=7truecm
\epsffile{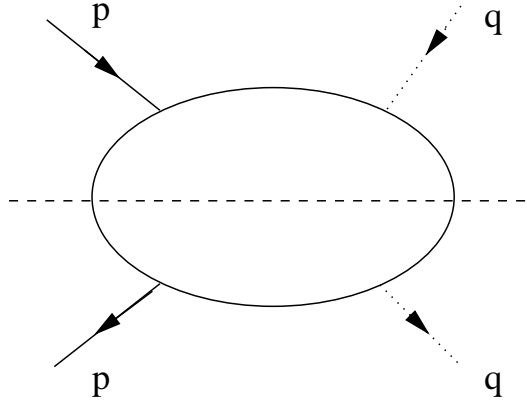}\\
\end{tabular}
\end{center}
\label{1}
\caption{{\em Deep inelastic structure function in $\pts$}}
\end{figure}

The leading contribution to the structure function
comes from the decomposition shown in Fig.2. Here, with the
notations of Ref. \cite{mue},
$\t$ is the hard part of the diagram,
i.e. the one in which
the large momentum flows, while $\l$ is the soft part. Decompositions with
more than two legs connecting the hard to the soft part are suppressed by
powers of $1/Q^2$.

\begin{figure}[htb]
\begin{center}
\begin{tabular}{c}
\epsfxsize=9truecm
\epsffile{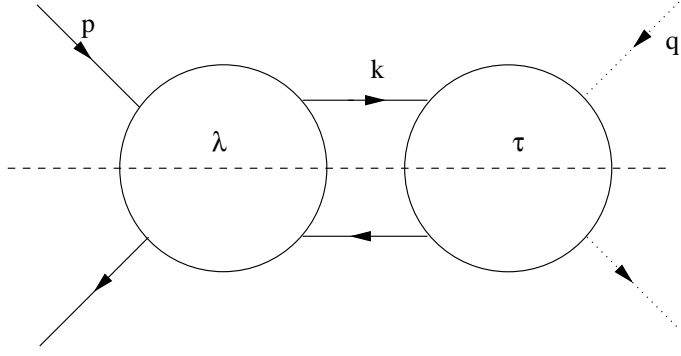}\\
\end{tabular}
\end{center}
\label{2}
\caption{{\em Relevant decomposition for the deep inelastic
structure function in $\pts$}}
\end{figure}

Such decomposition can be written in formulae as
\begin{equation}
\label{deco}
F(p,q)=\sum_\t \int V_\l(p,k) \hs H_\t(\kh,q)\hs \dk \; ,
\end{equation}
where
$V_\l(p,k)$ and $H_\t(k,q)$ are the discontinuities of the long
and short distance parts, respectively.

Moreover, in order to pick up the leading contribution in eq. (\ref{deco}),
the momentum $k$ which enters $\tau$ is taken to be collinear
to the external momentum $p$. Neglecting renormalization, let us
define for a given
decomposition
into a $\l$ and a $\t$ subdiagram
\begin{equation}
\label{cut}
v_\l(p^2,x)=\int V_\l(p,k)\hs x \hs\delta\left(x-\f{k_+}{p_+}\right)\hs\dk
\end{equation}
and
\begin{equation}
C_\tau(x,Q^2)=H_\tau(k^2=0,x,q^2).
\end{equation}
Here $v_\l(p^2,x)$ represents the contribution of $\l$ when the hard part
is contracted to a point, while $C_\tau(x,Q^2)$ is the hard part in which
one neglects the virtuality of the incoming momentum with respect to $Q^2$.
Since
\begin{equation}
x\simeq \f{Q^2}{2p_+q_-}
\end{equation}
and
\begin{equation}
H_\tau(\kh,q)=H_\tau(0,\f{Q^2}{2k_+q_-},q^2) ,
\end{equation}
using the definition of $\kh$ and eqs. (2.3)-(2.7) we can write
\begin{align}
\label{dis}
F(p,q)&=\sum_\t \int V_\l(p,k)\hs H_\t(\kh,q)\hs\dk\nn\\
&=\sum_\t\int V_\l(p,k)\hs\delta\left(u-\f{k_+}{p_+}\right)du
~C_\tau(x/u,Q^2)\hs \dk\nn\\
&=\sum_\t\int v_\l(p^2,u)\hs C_\t(x/u,Q^2)\hs\f{du}{u}\equiv \int
v(p^2,u)\hs C(x/u,Q^2)\hs \f{du}{u}.
\end{align}
The last integral defines the spacelike {\em cut vertex}
 $v(p^2,x)$
and  the corresponding coefficient function $C(x,Q^2)$.
As usual a simpler factorized expression for the structure function is
obtained by taking  moments with respect to $x$.
Defining the Mellin transform as
\begin{equation}
\label{Mellin}
f_\sigma=\int^1_0 dx \hs x^{\sigma-1} f(x)  ,
\end{equation}
we find immediately
\begin{equation}
\label{mcut}
F_\sigma(p^2,Q^2)=v_\sigma(p^2) \hs C_\sigma(Q^2).
\end{equation}

 It was shown in Refs.\cite{kounnas,mun} that the cut vertex represents
the analytic continuation in the spin variable of a matrix element
of operators of minimal twist.
This correspondence has been confirmed up to
two loops by direct calculation
of the anomalous dimensions of cut vertices and leading
twist operators \cite{kounnas}.
Hence, in the case of DIS, the factorized expression (\ref{mcut})
can be identified
with the one given by  OPE
\begin{equation}
\label{mope}
F_n(p^2,Q^2)=A_n(p^2) \hs C_n(Q^2)\;,
\end{equation}
where $A_n(p^2)$ are now matrix elements of local operators.
Thus, for integer values of $\sigma$, the coefficient function
which appears in (\ref{mcut}) is the same
as in (\ref{mope}). This fact will be used in the next section
where the evolution of the extended fracture function
will be shown to be
driven by the anomalous dimension of the same set of local operators.

\resection{A cut vertex approach to semi-inclusive processes}

Let us consider now, still within $\pts$, a deep inelastic reaction
in which a particle
with momentum
$p^\prime$ is inclusively observed in the final state, i.e. the process
\begin{equation}
p+J(q)\rightarrow p^\prime+X\nn.
\end{equation}
 By using the same
line of reasoning as for the inclusive case
we may define
a semi-inclusive structure function as (see Fig. 3)

\begin{figure}[htb]
\begin{center}
\begin{tabular}{c}
\epsfxsize=7truecm
\epsffile{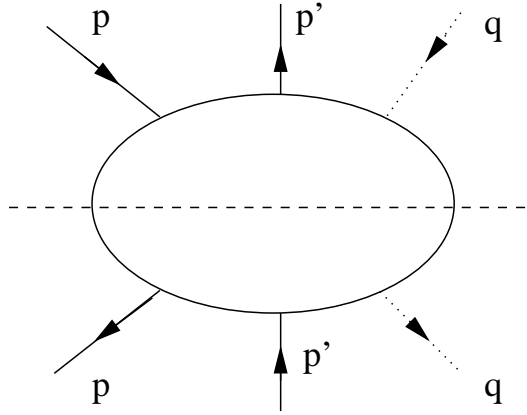}\\
\end{tabular}
\end{center}
\label{3}
\caption{{\em Deep inelastic semi-inclusive structure function in $\pts$}}
\end{figure}

\begin{equation}
W(p,p^\prime,q)=\f{Q^2}{2\pi} \sum_X\int d^6x~e^{iqx}
<p|J(x)|p^\prime X>
<X~p^\prime|J(0)|p>
\end{equation}
in terms of matrix elements of  the current operator between the incoming
hadron with momentum $p$ and the outgoing hadron with momentum
$p^\prime$ plus
anything.

When the observed particle has transverse momentum $p^{\prime 2}_\perp$
of order $Q^2$
the cross section is dominated by the current fragmentation mechanism
and  can be written in the usual
factorized way \cite{aemp}
\begin{equation}
\label{current}
W(p,p^\prime,q)=\int \f{dx^\prime}{x^\prime} \f{dz^\prime}{z^\prime}\hs
f_A(x^\prime,Q^2)\hs {\hat \sigma}(x/x^\prime,z^\prime,Q^2)
\hs D_{A^\prime}(z/z^\prime,Q^2) ,
\end{equation}
where
\begin{equation}
z=\f{pp^\prime}{pq}\simeq \f{p^\prime_-}{q_-}\;.
\end{equation}
In the language of cut vertices, eq.(\ref{current}) is a convolution  
of a spacelike and a timelike cut vertex
through a coefficient function \cite{gm}
\begin{equation}
W(p,p^\prime,q)=\int \f{dx^\prime}{x^\prime} \f{dz^\prime}{z^\prime}
v(p^2,x^\prime)\hs C(x/x^\prime,z^\prime,Q^2)
\hs v^\prime(p^{\prime 2},z/z^\prime).
\end{equation}

By contrast, the limit
$t=-(p-p^\prime)^2\ll Q^2$ is dominated by the target fragmentation
mechanism
and
has not been considered in either approach\footnote{In Ref. \cite{grau}
it has been shown at one loop
that in the limit $t\to 0$ a new collinear singularity
appears in the semi-inclusive cross section which cannot be absorbed
into parton densities and fragmentation functions, and so must be lumped
into a new phenomenological distribution, i.e. the fracture function.}.
Following the same steps as in the previous section,
one can argue that, in the region $t\ll Q^2$, the leading contribution to
the semi-inclusive cross section is given by the decomposition shown in
Fig.4.

\begin{figure}[htb]
\begin{center}
\begin{tabular}{c}
\epsfxsize=9truecm
\epsffile{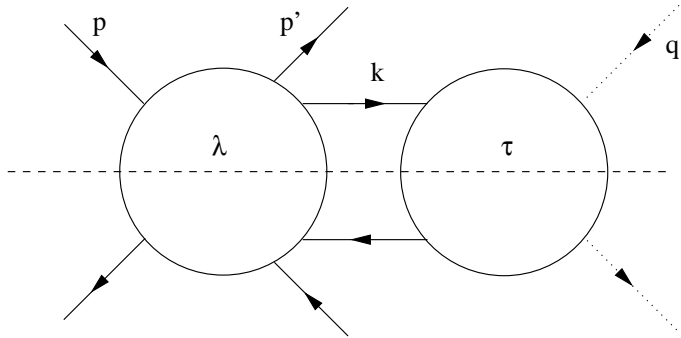}\\
\end{tabular}
\end{center}
\label{4}
\caption{{\em Relevant decomposition for the semi-inclusive
structure function in $\pts$ in the limit $t\ll Q^2$}}
\end{figure}

Such a decomposition implies
that an expansion similar to (\ref{dis}) holds, in terms of a new function
$v(p,p^\prime,\xb)$ and a coefficient function $C(\xb,Q^2)$
\begin{equation}
\label{semi}
W(p,p^\prime,q)=\int v(p,p^\prime,u)\hs C(\bar{x}/u,Q^2)\hs\f{du}{u}
\end{equation}
where we have defined a new variable $z$ as
\begin{equation}
z=\f{p^\prime q}{pq}\simeq \f{p^\prime_+}{p_+}
\end{equation}
and a rescaled variable $\bar{x}=x/(1-z)$. The new function
$v(p,p^\prime,\xb)$
is given by
\begin{equation}
\label{gencut}
v(p,p^\prime,\bar{x})=\int T(p,p^\prime,k)\hs\bar{x}\hs
\delta\left(\bar{x}-\f{k_+}{p_+-p^\prime_+}\right)\f{d^6k}{(2\pi)^6} ,
\end{equation}
where $T(p,p^\prime,k)$ is the discontinuity of a
six-point amplitude in the channel
$(p-p^\prime-k)^2$.
The function $v(p,p^\prime,\bar{x})$ is a new object that
we will call a {\em generalized cut vertex}, which depends both on $p$  
and $p^\prime$
and embodies all the leading mass singularities of the cross section.
By taking moments with respect to $\bar{x}$ as in eq.(\ref{Mellin}),
eq. (\ref{semi}) becomes
\begin{equation}
\label{semim}
W_\sigma(p,p^\prime,q)=v_\sigma(p,p^\prime)\hs C_\sigma(Q^2)
\end{equation}
that is a completely factorized expression analogous to (\ref{mcut}).

We are now going to show that this expansion holds
up to corrections suppressed by powers of $1/Q^2$.
In order to do so we will use the method
of infra-red power counting \cite{sterman,css} applied to our process.

\begin{figure}[htb]
\begin{center}
\begin{tabular}{c}
\epsfxsize=7truecm
\epsffile{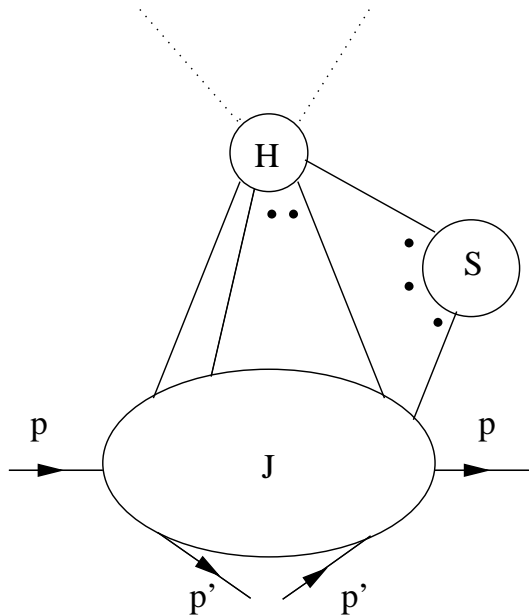}\\
\end{tabular}
\end{center}
\label{5}
\caption{{\em General reduced graphs which contribute to the semi-inclusive
structure function in $\pts$}}
\end{figure}

In order to get insight into the large $Q^2$ limit
of the semi-inclusive cross section, let us look at the singularities in
the limit $p^2$, $p^{\prime 2}$, $t\to 0$. The infra-red power counting
technique can predict the strength of such
singularities.
Starting from a given diagram, its {\em reduced} form in the large
$Q^2$ limit
is constructed by
simply contracting to a point all the lines whose momenta are not on shell.
The general reduced diagrams in the large $Q^2$ limit
for the process under study involve a jet subdiagram $J$,
composed by on-shell lines collinear to the incoming particle,
from which the detected particle emerges in the forward
direction\footnote{In the large $Q$ limit $p$ and $p^\prime$ can be
taken as parallel.},
a hard subgraph $H$ in which momenta
of order $Q$ circulate, which is connected to the jet by an arbitrary number
of collinear lines. Soft connections between $J$ and $H$
can be possibly collected into
a soft blob $S$ which is connected to the
rest of the diagram by an arbitrary number of lines (see Fig.5).
In $\pts$, by using power counting \cite{css},
we find\footnote{This fact has been verified by an explicit one-loop  
calculation in Ref. \cite{graz}.}
that the leading
contributions come from graphs with no soft lines and the minimum number of
collinear lines connecting the hard to the jet subdiagram, as in Fig.6.

\begin{figure}[htb]
\begin{center}
\begin{tabular}{c}
\epsfxsize=7truecm
\epsffile{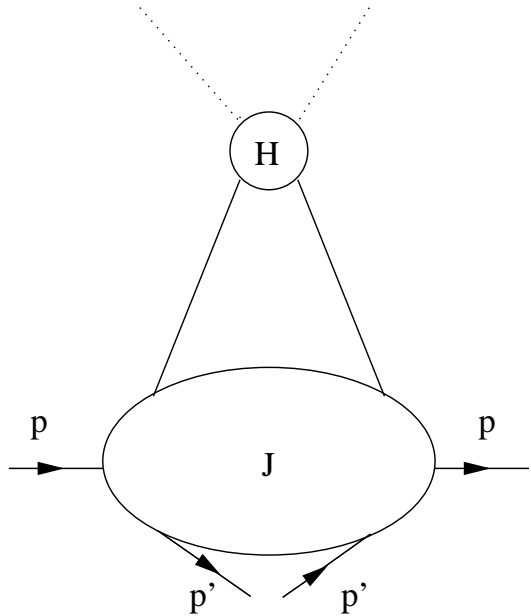}\\
\end{tabular}
\end{center}
\label{6}
\caption{{\em Leading contributions to the semi-inclusive
structure function in $\pts$}}
\end{figure}

Any other diagram containing additional collinear lines between
$J$ and $H$ is suppressed by powers of $1/Q^2$.
It follows that $W(p,p^\prime,q)$ is
of the following form
\begin{equation}
\label{fact}
W(p,p^\prime,q)=\int \f{d^6k}{(2\pi)^6}\hs T(p,p^\prime,k)\hs H({\hat k},q)
+{\cal O}(1/Q^2).
\end{equation}

It is now straightforward to show that eq. (\ref{fact}) is equivalent to
(\ref{semi}) with the substitution $H(0,x,q^2)=C(x,Q^2)$.
Thus the expansion (\ref{semi}) corresponds to
taking the leading part of the semi-inclusive cross section.

\resection{Extended fracture functions}

In the previous section we have given arguments for the validity of a  
generalized cut vertex expansion for the process $p+J(q)\to
p^\prime+X$ in the region
$t\ll Q^2$. We now want to investigate the consequences of such a result.

The coefficient function
which appears in (\ref{semi}) is the same as that of (\ref{dis})
since it comes
from the hard part of the graphs which is exactly the same as in DIS.
So we can draw the important conclusion that the
evolution of the coefficient function appearing in (\ref{semi}) is directly
related
to the anomalous dimension of the leading twist local operator which drives
the evolution of the DIS coefficient function.

Despite the fact that the theoretical framework in which we have been  
working
is the model field theory $\pts$, we expect that the main
consequences expressed
in (\ref{semi}) will remain valid also in a gauge theory
such as QCD.
The only further complications which are expected to arise
are due to soft gluon lines connecting the
hard to the
jet subdiagrams. Unlike in $\pts$, in QCD these diagrams are not
suppressed  by
power counting; hence
 the only way to get rid of such contributions is to show that they
cancel out. This is the issue of a complete factorization proof, which
is beyond the
aim of this paper. As already argued in Ref.\cite{soper} we do not
expect that this complication will destroy
factorization . In QCD,
by using renormalization group, we have
\begin{equation}
C^i_n(Q^2)\equiv C^i_n(Q^2/Q_0^2,\as)=
\left[ e^{\textstyle \int_{\as}^{\as(Q^2)} d\alpha
\f{\gamma^{(n)}(\alpha)}{\beta(\alpha)}}\right]_{ij}
~C_n^j(1,\as(Q^2)) ,
\end{equation}
where $Q_0$ is the renormalization scale, $\as\equiv \as(Q^2_0)$,
$\gamma^{(n)}$ is the anomalous dimension matrix of the relevant operators,
and an ordered exponential is to be understood.
Thus we can write the analogue of eq.(\ref{semim}) in QCD as
\begin{equation}
W_n(z,t,Q^2)=\sum_i {\cal M}^i_n(z,t,Q^2)~C_n^i (1,\as(Q^2))
\end{equation}
where, by following Ref. \cite{kounnas}, we have defined a  
$t$-dependent fracture function
\begin{equation}
\label{defi}
{\cal M}^j_n(z,t,Q^2)\equiv V^i_n(z,t,Q_0^2)
\left[ e^{\textstyle \int_{\as}^{\as(Q^2)} d\alpha
\f{\gamma^{(n)}(\alpha)}{\beta(\alpha)}}\right]_{ij}
\end{equation}
just in terms of a cut vertex $V^i_n(z,t,Q_0^2)$ (see Fig. 7).

\begin{figure}[htb]
\begin{center}
\begin{tabular}{c}
\epsfxsize=7truecm
\epsffile{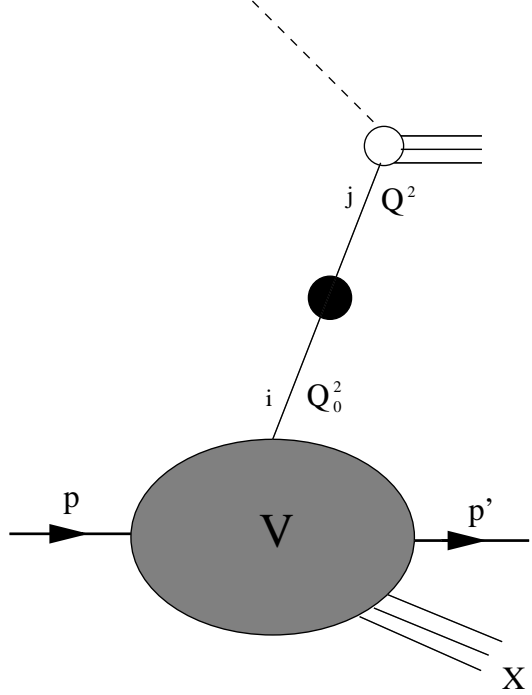}\\
\end{tabular}
\end{center}
\label{7}
\caption{{\em Extended fracture function}}
\end{figure}

Inverting the moments and expressing the extended fracture function
in terms of the usual Bjorken variable $x$,
one finds that ${\cal M}^i_{A,A'}(x,z,t,Q^2)$
obeys the simple homogeneous evolution equation
\begin{equation}
\label{eqomo}
Q^2\f{\d}{\d Q^2} {\cal M}^i_{A,A'}(x,z,t,Q^2)=
\sum_j\int_{\f{x}{1-z}}^1 \f{du}{u}K_{ij}(u,\as(Q^2))
\hs {\cal M}^j_{A,A'}(x/u,z,t,Q^2)
\end{equation}
where $K_{ij}(u,\alpha)$, defined as
\begin{equation}
K_{ij}(u,\alpha)\equiv \f{1}{2\pi i}
\int_{\f{1}{2}-i\infty}^{\f{1}{2}+i\infty} dn\hs
\gamma^{(n)}_{ij}(\alpha)\hs u^{-n} ,
\end{equation}
is the same DGLAP kernel which controls the evolution of the ordinary
parton distribution functions.
This result
looks particularly appealing since it means that the evolution
of the extended fracture function follows the usual perturbative behaviour.
One may ask at this point how this result matches with the peculiar equation
which drives the evolution of ordinary fracture functions \cite{frac}:
\begin{align}
\label{inomo}
Q^2 \f{\d}{\d Q^2}&M^j_{A,A'}(x,z,Q^2)=
\f{\as(Q^2)}{2\pi}
\int^1_{\f{x}{1-z}}\f{du}{u}\hs P_i^j(u) \hs M^i_{A,A'}(x/u,z,Q^2)\nn\\
&+ \f{\as (Q^2)}{2\pi}\int^{\f{x}{x+z}}_x \f{du}{x(1-u)}
\hs F_A^i(x/u,Q^2)\hs {\hat P}_i^{jl}(u)
\hs D_{l,A'}\left(\f{zu}{x(1-u)},Q^2\right).
\end{align}

The evolution equation for $M^j_{A,A^\prime}(x,z,Q^2)$ contains two
terms:
a homogeneous term describing the non-perturbative production
of a hadron coming from target fragmentation and an inhomogeneous term
whose origin is the perturbative fragmentation
due to initial state bremsstrahlung. As discussed in Ref. \cite{frac},
the separation between perturbative and non-perturbative
fragmentation introduces an arbitrary scale but the fracture function  
itself
$M^j_{A,A^\prime}(x,z,Q^2)$ does not depend on it.

It can be shown \cite{next}
that the  evolution equation (\ref{inomo}) can be
 derived
by an explicit calculation
using eq. (\ref{eqomo}) together with
Jet Calculus rules \cite{jet}. By defining in fact the ordinary fracture
function as an integral over $t$ up to a cut-off of
order $Q^2$, e.g. $\epsilon\hsp Q^2$ with $\epsilon < 1$:
\begin{equation}
M^j_{A,A^\prime}(x,z,Q^2)=\int^{\epsilon\hsp Q^2} dt\hs {\cal
M}^j_{A,A^\prime}(x,z,t,Q^2) ,
\end{equation}
the inhomogeneous term in the evolution equation is obtained by taking into
account the $Q^2$ dependence of the integration cut-off.

Moreover, the interplay between the scales $Q^2$ and $t$
has a sizeable effect in terms of
a new class of perturbative corrections of the form $\log Q^2/t$.
Such corrections are large and potentially dangerous
in the region $t\ll Q^2$ since they can ruin a reliable perturbative
expansion.
Those terms are naturally
resummed into eq. (\ref{defi}). For the
extended fracture function these corrections
do play an important role
for understanding the dynamics of semi-inclusive processes in the kinematic
region we have been considering here \cite{next}.

\resection{Conclusions}

In this paper we have presented an extension of the cut vertex
formalism to deal with semi-inclusive processes. We have shown that
an extended  fracture
function can be defined  in terms of a new kind of cut vertex.  Such  
an unintegrated, $t$-dependent
fracture function obeys a DGLAP evolution equation with the same
kernel
as the one
appearing in ordinary parton densities.
This result is particularly evident in the cut vertex approach
and agrees with the analysis made in a different context in
Ref.\cite{soper}. As a non-trivial check, the more complicated
evolution equation for ordinary fracture functions \cite{frac} can be  
obtained from the simpler one obeyed by extended fracture functions.

Although our results have been obtained within the $\pts$ model,
where a
treatment based on mass singularities and infra-red power counting
takes a much simpler form, we believe the extension to QCD to be
harmless. Indeed, the picture that emerges for semi-inclusive hadronic
processes in the target fragmentation region appears to be
fully consistent with the intuitive QCD-corrected parton picture.

In addition,
a new class of logarithmic contributions appears at $\Lambda^2 \ll t
\ll Q^2$.
Extended fracture functions
 could give an additional tool to handle this interesting regime
which certainly deserves further study \cite{next}.

We finally note that the relation between ordinary and extended
fracture functions is quite analogous to the one between the
experimentally defined/measured
 diffractive structure functions $F_2^{D(3)}(\xp,\beta,Q^2)$ and
$F_2^{D(4)}(\xp,\beta,Q^2,t)$ \cite{exp} which depend upon
observing large rapidity gaps. Relating such  quantities
to fracture functions is another interesting issue for future investigation.
\vspace{3mm}
\begin{center}
\begin{large}
{\bf Acknowledgements}
\end{large}
\end{center}
\noindent We would like to thank J.C. Collins, J. Kodaira and G.M. Shore
for helpful discussions.

\end{document}